\documentclass{aastex}
\usepackage{epsfig, natbib, graphicx,amsmath}

\def\wf{\textit{WFIRST}}

\begin{document}
\shortauthors{Patel, Millan-Gabet, Krist, Beichman}
\shorttitle{WFIRST CGI FITS Template}
\title{Standard FITS template for Simulated Astrophysical Scenes with the WFIRST Coronagraph}

\author{Rahul I. Patel\altaffilmark{1}, 
        Rafael Millan-Gabet\altaffilmark{1}, 
        John Krist\altaffilmark{2},
        Charles Beichman\altaffilmark{1,2,3}}
        
\altaffiltext{1}{Infrared Processing and Analysis Center, California Institute of Technology, Pasadena, CA 91125}
\altaffiltext{2}{Jet Propulsion Laboratory, California Institute of Technology, 4800 Oak Grove Dr. Pasadena 91109 }
\altaffiltext{3}{NASA Exoplanet Science Institute, California Institute of Technology, 770 S. Wilson Ave., Pasadena, CA 91125}


\begin{abstract}
    The science investigation teams (SITs) for the \wf\ coronagraphic instrument have begun studying the capabilities of the instrument to directly image reflected light off from exoplanets at contrasts down to contrasts of $\sim10^{-9}$ with respect to the stellar flux. Detection of point sources at these high contrasts requires yield estimates and detailed modeling of the image of the planetary system as it propagates through the telescope optics. While the SITs might generate custom astrophysical scenes, the integrated model, propagated through the internal speckle field, is typically done at JPL. In this white paper, we present a standard file format to ensure a single distribution system between those who produce the raw astrophysical scenes, and JPL modelers who incorporate those scenes into their optical modeling.  At its core, our custom file format uses FITS files, and incorporates standards on packaging astrophysical scenes. This includes spectral and astrometric information for planetary and stellar point sources, zodiacal light and extragalactic sources that may appear as contaminants. Adhering to such a uniform data distribution format is necessary, as it ensures seamless work flow between the SITs and modelers at JPL for the goals of understanding limits of the \wf\ coronagraphic instrument.

\end{abstract}

\section{Introduction}\label{sec:intro}

        The coronagraphic instrument (CGI) for the Wide-Field Infrared Survey Telescope (WFIRST) will be used to detect reflected starlight from exoplanets at planet-to-star contrasts of $\sim10^{-9}$. The WFIRST-CGI will have multiple observing modes. These consist of discovery modes using a Hybrid Lyot Coronagraph \citep[HLC;][]{Trauger2016} coupled with broadband optical imaging for detection of exoplanets and debris disks, and a Shaped Pupil Coronagraph \citep[SPC;][]{Carlotti2013,Riggs2014} and SPC disk mask \citep{Zimmerman2016} coupled with an integral field spectrograph (IFS) for characterization.         
        
        To properly characterize the performance capabilities of the WFIRST-CGI and improve upon current post-processing data reduction techniques, simulated coronagraphic data of selected targets must be analyzed. Typically, an astrophysical scene is generated, whereby information of the target star, astrometric and flux information for planets and background point sources, and other astrophysical noise is accumulated and propagated through a speckle field to model the light from the astrophysical scene moving through the internal telescope optics. In most cases, these two tasks are done separately. Thus, to ensure seamless work flow, a standard distribution system of the raw and processed astrophysical scene is necessary. 
        
        In the following article, we outline an efficient method of creating and distributing the scene information by using a multi-extension Flexible Image Transport System (FITS) file. We begin in \S~\ref{sec:datatype} by describing each simulation component and the structure of the multi-extension FITS file. We then describe the format and required header keywords for the primary extension in \S~\ref{sec:primary_header}. Next, we describe the details for constructing the extensions that must hold table formatted information, such as fluxes and astrmetric information of point sources, in \S~\ref{sec:tabledata}, . Lastly we describe the details for constructing the extensions that hold the spectral data cubes of the exozodiacal emission and background galactic contaminants in \S~\ref{sec:imagecubes}.

\section{Supported Data \& FITS Adoption}\label{sec:datatype}
    
    A typical astrophysical scene will have multiple independent components, each described by a myriad of meta-data, simulated to the specifications of different observing modes, at varying complexities. Regardless of the instrumental dependencies, each scene will be created using a combination of the following components:
    
    \begin{itemize}
        \item Astrometric information of generated point sources (background stars and planets).
        \item Contrast and flux information of the central star and point sources.
        \item Spectral data cubes of extragalactic background sources.
        \item Spectral data cubes of exozodiacal emission.
    \end{itemize}
    
    \noindent It is important to encode the individual elements of the astrophysical scene, thereby providing transparency and clarity. To encode each individual aspect of the astrophysical scene, we decided to package the data into a FITS file, each with their individual extension, encompassed as a single FITS file. In addition to the standard FITS header keywords (e.g., \texttt{BITPIX}, \texttt{NAXIS}, \texttt{EXTEND}, etc.), we outline the format for each section and allowed meta-data header information in the following sections.

    \subsection{General Format}

    Each simulation package will be a multi-extension FITS formatted file. A single FITS file will be unique to a single star and observation epoch, and include at the very least the primary extension described in \S~\ref{sec:primary_header}. Any additional extensions are optional and are not subject to being placed in any particular order. The extensions are however subject to standard names, and the user may access each by calling on the extension \textit{name}. A list of acceptable extensions are listed in Table~\ref{tab:full_file}, with additional information in the succeeding sections.
    
    \begin{deluxetable}{lll}
    \tabletypesize{\normalsize}
    \tablecaption{All Extension Labels and Descriptions.}\label{tab:full_file}
    \tablehead{ \colhead{Extension} & \colhead{Extension} & \colhead{Description}\\
                 \colhead{Name}     & \colhead{Type}      & \colhead{} }
    \startdata
    \texttt{PRIMARY}        & PrimaryHDU  & Primary header: Contains stellar and simulation parameters in header.\\
    \texttt{CONTRAST\_BB}    & BinTableHDU & Broadband contrast and fluxes\\
    \texttt{CONTRAST\_IFS}  & BinTableHDU & IFS contrast and fluxes\\
    \texttt{ASTROMETRY}     & BinTableHDU & Positional information of point sources\\
    \texttt{ZODI\_BB}       & ImageHDU    & Broadband exozodiacal image cube\\
    \texttt{ZODI\_IFS}      & ImageHDU    & IFS exozodiacal image cube\\
    \texttt{BACKGROUND\_BB} & ImageHDU    & Broadband extragalactic image cube\\
    \texttt{BACKGROUND\_IFS}& ImageHDU    & IFS extragalactic image cube
    \enddata
    \tablecomments{\small General description of all possible extensions that may be included in the multi-extension FITS file. All extensions, except the \texttt{PRIMARY} extension, are optional.}
    \end{deluxetable}

\section{Standard Primary Header}\label{sec:primary_header}
    
    The primary header of the FITS file is reserved to host information for the primary star. The primary header will not contain any image or table, but holds meta-data about the observed target in the headers. A required list of header keywords and associated data types are listed in Table~\ref{tab:primary_header}.
    
    \begin{deluxetable}{llll}
    \tabletypesize{\Large}
    \tablecaption{Primary Header FITS Keywords\label{tab:primary_header}}
    \tablehead{\colhead{Keyword} & \colhead{Data} &\colhead{Fixed}& \colhead{Description}\\
                \colhead{}       &\colhead{Type}  &\colhead{Value} &\colhead{}
                }
    \startdata
\texttt{SIMDATE}  & string (10-char) & \nodata & Date of simulation YYYY-MM-DD \\
\texttt{EPOCH}    & float            & \nodata & UT date of observation \\
\texttt{OBJNAME}  & string           & \nodata & Target's SIMBAD ID \\
\texttt{RA\_OBJ}  & float            & \nodata & Right Ascension (deg)\\
\texttt{DEC\_OBJ} & float            & \nodata & Declination (deg)\\
\texttt{PMRA}     & float            & \nodata & RA proper motion (mas/yr) \\
\texttt{PMDEC}    & float            & \nodata & DEC proper motion (mas/yr)\\
\texttt{PLX}      & float            & \nodata & Parallax (mas) \\
\texttt{SPT}      & string           & \nodata & Spectral Type\\
\texttt{V\_OBJ}   & float            & \nodata & V band (mag)\\
\texttt{B\_OBJ}   & float            & \nodata & B band (mag) \\
\texttt{J\_OBJ}   & float            & \nodata & J band (mag)\\
\texttt{H\_OBJ}   & float            & \nodata & H band (mag)\\
\texttt{KS\_OBJ}  & float            & \nodata & Ks band (mag)\\
\texttt{COMMENT}  & string           & \nodata & Comments related to simulation.
    \enddata
    \tablecomments{Standard keywords in the \texttt{PRIMARY} extension. All these headers are required, and keywords cannot be changed.}
    \end{deluxetable}

\section{Table Data}\label{sec:tabledata}

    \subsection{Astrometry}\label{sec:astrometry}
    
    Point need to have their astrometric information included in the \texttt{ASTROMETRY} extension. Instead of absolute coordinate information, coordinate offsets from the stellar center must be listed. The data must be encoded in a FITS binary table format. Columns are designated for each point source, and a maximum of two rows for each column are designated for offsets in right ascension and declination for the first and second row, respectively. 
    
    The designation for each point source (i.e., column title) is encoded in the \texttt{TTYPEi} keyword, where \texttt{i} is the $i^\text{th}$ column. The column header for simulated planets can be listed as ``Planet \textit{j} Offset'', where \textit{j} is the $j^\text{th}$ simulated planet. Background field stars can be listed as ``Field Star \textit{k} Offset'', where \textit{k} is the $k^\text{th}$ simulated field star. Neither \textit{j} or \textit{k} need to be equal to \texttt{i}. In addition, the column header can also be given custom titles, if the user wishes. For instance, planets can also be described by their official designation and/or additional model descriptions (e.g., ``47UMa\_c\_model3.2''). The astrometric offsets should be in units of ``arcseconds'', listed under the FITS keyword \texttt{TUNITi}. 
    
    \begin{deluxetable}{llcl}
    \tablewidth{0pt}
    \tabletypesize{\large}
    \tablecaption{Astrometry extension header keywords.\label{tab:astrometry}}
    \tablehead{\colhead{Keyword} & \colhead{Data} &\colhead{Fixed}& \colhead{Description}\\
                \colhead{}       &\colhead{Type}  &\colhead{Value} &\colhead{}
                }
    
    \startdata
    \texttt{EXTNAME}  & string & ASTROMETRY & extension name \\
    \texttt{TFIELDS}  & int    & \nodata    & number of table fields\\
    \texttt{FIELDCOL} & int    & \nodata    & Starting column of field star flux \\
    \texttt{PLCOL}    & int    & \nodata    & Starting column of planet contrasts  \\
    \texttt{TTYPEi}   & string & Planet \textit{j} Offset or Field Star \textit{k} Offset & Column Name.\\
    \texttt{TFORMi}   & string & E    & Column Data Type\\
    \texttt{TUNITi}   & string & arcsec    & Column Units\\
    \texttt{COMMENT}  & string & \nodata    & Comments related to simulation.
    \enddata
    \tablecomments{Standard keywords for the extension which contains the point-source astrometry.}
    \end{deluxetable}

    \subsection{Contrast \& Fluxes}\label{sec:contrastfluxes}
    
    The flux of each point source in \S~\ref{sec:astrometry} and the target star must be tabulated in the binary table \texttt{CONTRAST\_BB} extension for broadband wavelengths and$/$or the \texttt{CONTRAST\_IFS} extension, for IFS wavelengths. Columns are designated for each source, while each row corresponds to a different wavelength channel for the flux in each ``cell''. In the case of simulated planets, the fluxes must be listed as a flux ratio, relative to the central star. The target star and any background field stars should have absolute fluxes in units of ``Jy.'' To simplify the modelling, the fluxes of the central star and background stars \textit{must} be obtained using a flat response function over the listed bandpass be flat.
    
    The number of rows is determined by the number of spectral channels the user wishes to encode. For broadband fluxes and contrasts, a maximum of four rows --- one for each of the WFIRST-CGI broadband channels --- can be added to the binary table. Broadband flux information must be included in the \texttt{CONTRAST\_BB} extension only. The accompanying central wavelength and bandwidth must also be included into the header keywords \texttt{CRWAVEi} and \texttt{BWWAVEi}, respectively. Here, \texttt{i} $\in \{1,2,3,4\}$, to denote the four WFIRST-CGI imager wavelength channels. Point sources simulated at the IFS channels will have fluxes calculated at a larger number of wavelengths and included in the \texttt{CONTRAST\_IFS} extension only. For the latter, instead of listing each wavelength in the header, we require that the wavelength range and resolution be listed, from which the wavelength for each row can be calculated.
    
    For both extensions, the first column will always include the target spectrum, indicated by the header keyword \texttt{STARCOL}=1. The number of simulated field stars/planets must also be indicated in the header, as well as the starting column for each type of point source. The combination of the number of sources, starting column, and column title will allow the user to determine where each point source spectrum is encoded. As described in \S~\ref{sec:astrometry}, point source columns can be designated using a generic sequential name under the \texttt{TTYPE} keyword, as shown in Tables~\ref{tab:bbflux} and \ref{tab:ifsflux}, or with a custom title. The full list of IPAC header keywords for the broadband and IFS flux extensions can be found in Tables~\ref{tab:bbflux} and \ref{tab:ifsflux}.
    
    \begin{deluxetable}{llcl}
    \tablewidth{0pt}
    \tabletypesize{\normalsize}
    \tablecaption{Broadband contrast extension header keywords.\label{tab:bbflux}}
    \tablehead{\colhead{Keyword} & \colhead{Data} &\colhead{Fixed}& \colhead{Description}\\
                \colhead{}       &\colhead{Type}  &\colhead{Value} &\colhead{}
                }
    
    \startdata
\texttt{EXTNAME}  & string & CONTRAST\_BB & extension name \\
\texttt{CRWAVEi}  & int    & \nodata     & Broadband channel \texttt{i} center wavelength (nm)\\
\texttt{BWWAVEi}  & int    & \nodata    &  Broadband channel \texttt{i} bandwidth (\%) \\
\texttt{STARCOL}  & int    & 1       &  Begin Stellar Flux Column\\
\texttt{FIELDCOL} & string & \nodata &  Begin Field Star Flux Column\\
\texttt{PLCOL}    & string & \nodata &  Begin Planet Contrast Column\\
\texttt{NPL}      & string & \nodata & Number of Planets \\
\texttt{NFIELD}   & string & \nodata & Number of Field Stars\\
\texttt{TTYPEi}   & string & Stellar Flux/Field Star \texttt{i} Flux/Planet \texttt{i} Contrast & Column Name\\
\texttt{TFORMi}   & string & E    & Column Data Type\\
\texttt{TUNITi}   & string & Jy or CONTRAST  & Column Units\\
\texttt{COMMENT}  & string & \nodata    & 
    \enddata
    \tablecomments{Standard keywords for the extension which contains contrast and flux at broadband detector channels. The stellar flux must be the first column in the table. The column name depends on whether it is flux for the central star, background stars or contrast for simulated planets.}
    \end{deluxetable}
    
   \begin{deluxetable}{llcl}
    \tablewidth{0pt}
    \tabletypesize{\normalsize}
    \tablecaption{IFS contrast extension header keywords.\label{tab:ifsflux}}
    \tablehead{\colhead{Keyword} & \colhead{Data} &\colhead{Fixed}& \colhead{Description}\\
                \colhead{}       &\colhead{Type}  &\colhead{Value} &\colhead{}
                }
    
    \startdata
\texttt{EXTNAME}  & string & CONTRAST\_IFS & extension name \\
\texttt{MINWAVE}  & int    & \nodata &  Blue end of IFS (nm) \\
\texttt{MAXWAVE}  & int    & \nodata &  Red end of IFS (nm) \\
\texttt{NWAVE}    & int    & \nodata &  Number of IFS slices \\
\texttt{STARCOL}  & int    & 1       &  Begin Stellar Flux Column\\
\texttt{FIELDCOL} & string & \nodata &  Begin Field Star Flux Column\\
\texttt{PLCOL}    & string & \nodata &  Begin Planet Contrast Column\\
\texttt{NPL}      & string & \nodata &  Number of Planets \\
\texttt{NFIELD}   & string & \nodata &  Number of Field Stars\\
\texttt{TTYPEi}   & string & Stellar Flux or Field Star \texttt{i} Flux or Planet \texttt{i} Contrast & Column Name\\
\texttt{TFORMi}   & string & E    & Column Data Type\\
\texttt{TUNITi}   & string & Jy or CONTRAST  & Column Units\\
\texttt{COMMENT}  & string & \nodata    &  \nodata
    \enddata
    \tablecomments{\small Standard keywords for the extension which contains contrast and flux at IFS detector channels. The stellar flux must be the first column in the table. The column name depends on whether it is flux for the central star, background stars or contrast for simulated planets.}
    \end{deluxetable}

\section{Spectral Image Cubes}\label{sec:imagecubes}

    Spectral data cubes will be used to encode exozodiacal and background galactic emission. These images should not include instrumental effects. In other words, the flux from the background galaxies and exozodiacal light should be encoded in the spectral cube as raw astrophysical emission, prior to their light entering the telescope. If the user wishes to include WCS data, they may do so by including the WCS header keywords in each extension (\texttt{WCSAXES, CRPIX1, CRPIX2, CDELT2, CDELT2, CTYPE1, CTYPE2, CRVAL1, CRVAL2, LATPOLE}). However, these are not mandatory.
    
    \subsection{Exozodiacal Images}\label{sec:exozodii}
    
    Simulated exozodical images are to be included in the \texttt{ZODI\_BB} and \texttt{ZODI\_IFS} extensions, depending on whether the images were generated at broadband wavelengths or at finer IFS channels. The images should be packaged as image cubes of size $m \times m \times N$, where $m$ is the number of pixels in a single slice and $N$ is size of the spectral dimension.
    
    Similar to the guidelines in \S~\ref{sec:contrastfluxes}, images at broadband wavelengths must be specified using the \texttt{CRWAVEi} and \texttt{BWWAVEi} header keywords, while IFS generated images should include the wavelength range and resolution, as described in Tables~\ref{tab:zodibb} and \ref{tab:zodiifs}. The pixel intensity must be in absolute flux units of ``Jy''. This must also be reflected in the standard header keyword \texttt{BUNIT} = ``Jy''.

   \begin{deluxetable}{llcl}
    \tablewidth{0pt}
    \tablecaption{Broadband exozodiacal image-cube extension header keywords.\label{tab:zodibb}}
    \tablehead{\colhead{Keyword} & \colhead{Data} &\colhead{Fixed}& \colhead{Description}\\
                \colhead{}       &\colhead{Type}  &\colhead{Value} &\colhead{}
                }
    \startdata
\texttt{EXTNAME}  & string & ZODI\_BB & extension name \\
\texttt{OBSMODE} & string & BB & Broadband setting\\
\texttt{CRWAVEi}  & int    & \nodata     & Broadband channel \texttt{i} center wavelength (nm)\\
\texttt{BWWAVEi}  & int    & \nodata    &  Broadband channel \texttt{i} bandwidth (\%) \\
\texttt{BUNIT}    & string & Jy   &   Flux Units\\
\texttt{COMMENT}  & string & \nodata   &  \nodata
    \enddata
    \tablecomments{Standard keywords for the extension which contains the simulated exozodical image cube at broadband detector channels.}
    \end{deluxetable}

   \begin{deluxetable}{llcl}
    \tablewidth{0pt}
    \tabletypesize{\large}
    \tablecaption{IFS exozodiacal image-cube extension header keywords.\label{tab:zodiifs}}
    \tablehead{\colhead{Keyword} & \colhead{Data} &\colhead{Fixed}& \colhead{Description}\\
                \colhead{}       &\colhead{Type}  &\colhead{Value} &\colhead{}
                }
    
    \startdata
\texttt{EXTNAME}  & string & ZODI\_IFS & extension name \\
\texttt{OBSMODE} & string & IFS & IFS setting\\
\texttt{MINWAVE}  & int    & \nodata   & Blue end of IFS (nm) \\
\texttt{MAXWAVE}  & int    & \nodata   &  Red end of IFS (nm) \\
\texttt{NWAVE}    & int    & \nodata   & Number of IFS slices \\
\texttt{BUNIT}    & string & Jy        &  Flux Units\\
\texttt{COMMENT}  & string & \nodata   & \nodata
    \enddata
    \tablecomments{Standard keywords for the extension which contains the simulated exozodiacal image cube at the IFS detector channels.}
    \end{deluxetable}
    
    \subsection{Extragalactic Background Images}

    Data cubes containing simulated extragalactic background images should be packaged in the \texttt{BACKGROUND\_BB} and \texttt{BACKGROUND\_IFS} extensions, in a similar format to the exozodiacal data cubes in \S~\ref{sec:exozodii}. Similarly, the data cubes will have dimensions of $m \times m \times N$.

   \begin{deluxetable}{llcl}
    \tablewidth{0pt}
    \tabletypesize{\large}
    \tablecaption{Broadband Background extragalactic image-cube extension header keywords.\label{tab:EGBbb}}
    \tablehead{\colhead{Keyword} & \colhead{Data} &\colhead{Fixed}& \colhead{Description}\\
                \colhead{}       &\colhead{Type}  &\colhead{Value} &\colhead{}
                }
    
    \startdata
\texttt{EXTNAME}  & string & BACKGROUND\_BB & extension name \\
\texttt{OBSMODE} & string & BB & Broadband setting\\
\texttt{CRWAVEi}  & int    & \nodata     & Broadband channel \texttt{i} center wavelength (nm)\\
\texttt{BWWAVEi}  & int    & \nodata    &  Broadband channel \texttt{i} bandwidth (\%) \\
\texttt{BUNIT}    & string & Jy         &  Flux Units\\
\texttt{COMMENT}  & string & \nodata    & \nodata
    \enddata
    \tablecomments{Standard keywords for the extension which contains the simulated extragalactic background image cube at the broadband detector channels.}
    \end{deluxetable}

   \begin{deluxetable}{llcl}
    \tablewidth{0pt}
    \tabletypesize{\large}
    \tablecaption{IFS Background extragalactic image cube extension header keywords.\label{tab:EGBifs}}
    \tablehead{\colhead{Keyword} & \colhead{Data} &\colhead{Fixed}& \colhead{Description}\\
                \colhead{}       &\colhead{Type}  &\colhead{Value} &\colhead{}
                }
    
    \startdata
\texttt{EXTNAME}  & string & BACKGROUND\_IFS & extension name \\
\texttt{OBSMODE} & string & IFS & IFS setting\\
\texttt{MINWAVE}  & int    & \nodata   & Blue end of IFS (nm) \\
\texttt{MAXWAVE}  & int    & \nodata   &  Red end of IFS (nm) \\
\texttt{NWAVE}    & int    & \nodata   & Number of IFS slices \\
\texttt{BUNIT}    & string & Jy   &  Begin Field Star Flux Column\\
\texttt{COMMENT}  & string & \nodata   & \nodata
    \enddata
    \tablecomments{Standard keywords for the extension which contains the simulated extragalactic image cube at the IFS detector channels.}
    \end{deluxetable}
    
\section{Conclusion}

    Our goal is to provide a standard file system for packaging of raw astrophysical scenes, simulated for the WFIRST-CGI. Since the modelers will need to incorporate different components in the scene, we present a multi-extension FITS format to allow for efficient data sharing between modelers and SIT teams. 

\acknowledgments This research was carried out at the Jet Propulsion Laboratory, California Institute of Technology, under a contract with the National Aeronautics and Space Administration. 


\end{document}